
\input harvmac.tex
\input tables.tex
\def\KB{\overline{K}^0}
\def\PT{\widetilde P}
\def\singlespace{\baselineskip 12 pt}

\hbox to 16.5truecm{\hfil ROMA Preprint n. 1070-1994}
\hbox to 16.5truecm{\hfil NAPOLI \ \qquad  DSF-T-56/94}
\hbox to 16.5truecm{\hfil hep-ph/9411286}
\vskip 1.5 cm
\centerline{\titlerm {Nonleptonic weak decays of charmed mesons}}
\vskip 1.5 cm
\centerline{F. Buccella$^{(a)}$, M. Lusignoli$^{(b,*)}$, G. Miele$^{(c)}$,
 A. Pugliese$^{(b,*)}$ and P. Santorelli$^{(c)}$}
\vskip 20pt
\centerline{
$^{(a)}$ {\it Dipartimento di Scienze Fisiche, Universit\`a di Napoli,
Napoli, Italy}}
\centerline{
$^{(b)}$ {\it Dipartimento di Fisica, Universit\`a ``La Sapienza'', Roma,}}
\centerline{{\it and INFN, Sezione di Roma I, Roma, Italy}}
\centerline{
$^{(c)}$ {\it Dipartimento di Scienze Fisiche, Universit\`a di Napoli,}}
\centerline{{\it and INFN, Sezione di Napoli, Napoli, Italy}}
\vskip 3 truecm
\centerline {ABSTRACT}
\vskip 10truept
\singlespace\noindent
A previous analysis of two-body Cabibbo allowed nonleptonic decays of
$D^0$ mesons and of Cabibbo allowed and first-forbidden decays of
$D^+$ and $D_s^+$ has been adjourned using more recent
experimental data and extended to the Cabibbo forbidden decays of $D^0$.
Annihilation and W-exchange contributions as well as final state
interaction effects (assumed to be dominated by nearby resonances)
have been included and are in fact crucial to obtain a
reasonable agreement with the experimental data, which show large
flavour SU(3) violations.
New fitting parameters are necessary
to describe rescattering effects for Cabibbo forbidden $D^0$
decays, given the lack of experimental informations on isoscalar
resonances. We
keep their number to a minimum - three - using phenomenologically based
considerations. We also discuss CP violating asymmetries.

\vskip 1 cm
\noindent
$^{(*)}$ partially supported by the European Community under the Human
Capital and Mobility Programme, contract CHRX-CT93-0132.

\vfill\eject
\baselineskip 18 truept
\parindent=1cm
\noindent
This paper replaces the previous one with the same title submitted to
the ICHEP94 conference (ref. gls0658), also circulated as ROMA preprint
n.1026-1994 / NAPOLI DSF-T-18/94. The 1994 edition of the Particle Data
(Phys. Rev. D 50 (August 1994) part I) did in fact show
noticeable changes in the experimental branching ratios for D decays. We
therefore made a new analysis, that we report here.

\newsec{Introduction}
A theoretical description of exclusive nonleptonic decays of charmed hadrons
based on general principles is not yet possible. Even if the short distance
effects due to hard gluon exchange can be resummed and an effective
hamiltonian has been constructed (recently, at next-to-leading order
\ref\rHEFNL{G. Altarelli, G. Curci, G. Martinelli and S. Petrarca,
Nucl. Phys. B 187 (1981) 461 \semi A.J. Buras, M. Jamin, M.E. Lautenbacher
and P.E. Weisz,
 Nucl.Phys. B 370 (1992) 69 and Nucl.Phys. B 375 (1992) 501 (addendum)
\semi M. Ciuchini, E. Franco, G. Martinelli and L. Reina,
Nucl. Phys. B 415 (1994) 403.}
), the evaluation of its matrix elements requires
nonperturbative techniques. Waiting for future progress in lattice QCD
calculations one has to rely on approximate methods and/or models.

The largely different lifetimes of charmed hadrons make it clear that the
infinitely heavy quark limit is quite far from the actual situation.
Therefore, the expansion in inverse powers of the heavy quark mass
characteristic of heavy quark effective theory
\ref\rHQET{see M. Neubert, Phys. Reports 245 (1994) 259.}
is presumably not a useful tool in this case. Moreover, the methods of
HQET are not obviously extended to cope with exclusive hadronic decays.
On the other hand, the simple factorized ansatz for the matrix elements
is known not to describe properly Cabibbo allowed $D^0$ decays. The
color-suppression of some contributions seems in fact to be stronger
than the factor 1/3 expected from QCD
\ref\rBSW{M. Bauer and B. Stech,
Phys.Lett. 152 B (1985) 380 \semi M. Bauer, B. Stech and M. Wirbel,
Zeits.f.Phys. C 34 (1987) 103 \semi
A.J. Buras, J.-M. G\'erard and R. R\"uckl, Nucl.Phys.
B 268 (1986) 16.}
and the data exhibit large phase differences between amplitudes
with definite isospin.
We are thus forced, still using the factorization approximation as a
starting point of the matrix element evaluation,
to include important corrections due to rescattering effects in the
final states. This we do assuming the dominance of nearby resonances
and taking from experiment - when possible - their masses and widths.
We also include W-exchange and annihilation
contributions that turn out to be larger than generally believed.
The presence of nearby resonances may well have the effect of increasing
these terms relative to their naive PCAC estimates.

In two previous papers the Cabibbo allowed
\ref\rBUCCA{F. Buccella, M. Lusignoli, G. Miele and A. Pugliese,
Zeits.f.Phys. C 55 (1992) 243.}
 two-body decays of charmed mesons were described
in the framework discussed above and the model was applied
to the analysis of Cabibbo first-forbidden decays of the charged mesons
\ref\rBUCCB{F. Buccella, M. Lusignoli, G. Mangano, G. Miele,
A. Pugliese and P. Santorelli, Phys.Lett. B 302 (1993) 319.}
{}~$D^+$ and $D_s^+$ and to their CP violating asymmetries.

The considerable success of that analysis prompts us to extend it
to the Cabibbo forbidden
two-body decays of $D^0$. The recent experimental determination of the
branching ratio B.R.($D^+ \to \pi^+\pi^0$) = 0.25$\pm$0.07 \%
\ref\rPDG{Review of Particle Properties, Particle Data Group,
Phys. Rev. D50 (1994) part I.}
\ref\rCLEODP{CLEO Collaboration: M. Selen et al.,
Phys. Rev. Lett. 71 (1993) 1973.}
- that agrees with our prediction \rBUCCB ~- allows to perform an
amplitude analysis on the
complex of $D \to \pi\pi$ decays, that shows a large ($\simeq 90^{\circ}$)
phase difference between $I=0$ and $I=2$ amplitudes. Moreover,
a comparison of the $I=2$ amplitude with the $I={3 \over 2}$
from $D^+ \to \pi^+ \KB$ shows a considerable violation of
flavour SU(3) in the direction of larger $\pi\pi$ amplitudes;
on the other hand, it is known since a long time and recently
confirmed \rCLEODP ~that the ratio of $D^0$ decay branching fractions
to $K^+ K^-$ and to $\pi^+ \pi^-$ is much {\it larger} than the SU(3)
prediction (i.e. 1), showing an opposite pattern of SU(3) breaking
in exotic and nonexotic channels. Another striking signal of the
importance of SU(3) violations is given by the value, quite similar
to other Cabibbo forbidden decays, of the B.R.($D^0 \to \KB K^0$)
that should be vanishing in the symmetric limit.

Our model describes satisfactorily the experimental situation. For
what concerns SU(3) breaking in exotic channels
it is the combination of several small effects that yields a large
result. These effects would not be enough to explain the large
$K^+ K^-$ to $\pi^+ \pi^-$ ratio and to produce a nonvanishing
decay rate to $\KB K^0$:
in the nonexotic channels the rescattering effects are essential.

In order to introduce these rescattering effects
 we need to know masses, widths and couplings of
yet unobserved spinless, isoscalar resonances with positive
and negative parity and masses around 1.9 GeV. One expects
for each parity two
resonances of this type, a SU(3) singlet and a member
of an octet, that generally mix among themselves.
Such a large number of
new parameters to fit eight new data (or limits) for branching ratios
is obviously unappealing, unless some arguments can be given to reduce it.
In the following we will show that reasonable phenomenological
assumptions may reduce the number of new parameters to three.

We have to determine these by a fit to the data.
Before doing that we repeated the fit
to all Cabibbo allowed and to charged meson first-forbidden decay
branching ratios, which in the meantime have got lower error bars and
in some cases have also changed. The model is therefore passing a more
demanding test.

\newsec{Decay amplitudes in the factorized approximation}
The effective weak hamiltonian for Cabibbo allowed nonleptonic decays of
charmed particles is given by ($U_{ij}$ are elements of the CKM matrix)
\eqn\eHEFFCA{\eqalign{ H_{eff}^{\Delta C = \Delta S} =
       {G_F \over \sqrt 2}\,U_{ud}\,U_{cs}^*\;
      \bigl[&C_2\;\bar{s}^\alpha\,\gamma_\mu\,(1-\gamma_5)\,c_\alpha\;
       \bar{u}^\beta\,\gamma^\mu\,(1-\gamma_5)\,d_\beta + \cr
  &      C_1\;\bar{u}^\alpha\,\gamma_\mu\,(1-\gamma_5)\,c_\alpha\;
       \bar{s}^\beta\,\gamma^\mu\,(1-\gamma_5)\,d_\beta \bigr] +
      {\rm h.c.}\ ,\cr}}
while for
$\Delta C = -\Delta S$ processes the hamiltonian is obtained from
the same equation with the substitution $s \leftrightarrow d$.
The effective weak hamiltonian for Cabibbo first-forbidden nonleptonic
decays is
\eqn\eHEFFCF{\eqalign{ H_{eff}^{\Delta C =\pm 1, \Delta S=0} &=\;
       {G_F \over \sqrt 2}\,U_{ud}\,U_{cd}^*\;
      \bigl[C_1\;Q_1^d\;+\;C_2\;Q_2^d \bigr]\; +\cr
    &+\;   {G_F \over \sqrt 2}\,U_{us}\,U_{cs}^*\;
      \bigl[C_1\;Q_1^s\;+\;C_2\;Q_2^s \bigr]\; -\cr
    &-\;   {G_F \over \sqrt 2}\,U_{ub}\,U_{cb}^*
      \; \sum_{i=3}^6 \;C_i\; Q_i \;\;\;+\;\; {\rm h.c.}\,,\cr}}
In eq. \eHEFFCF ~the operators are \ref\rGILMAN{F.J. Gilman and M.B. Wise,
Phys.Rev. D 20 (1979) 2392.}:
\eqn\eOPER{\eqalign{
Q_1^d =&\;
       \bar{u}^\alpha\,\gamma_\mu\,(1-\gamma_5)\,d_\beta\;
       \bar{d}^\beta\,\gamma^\mu\,(1-\gamma_5)\,c_\alpha, \cr
Q_2^d =&\;
       \bar{u}^\alpha\,\gamma_\mu\,(1-\gamma_5)\,d_\alpha\;
       \bar{d}^\beta\,\gamma^\mu\,(1-\gamma_5)\,c_\beta, \cr
Q_3 =&\;
       \bar{u}^\alpha\,\gamma_\mu\,(1-\gamma_5)\,c_\alpha\;
      \sum_q\; \bar{q}^\beta\,\gamma^\mu\,(1-\gamma_5)\,q_\beta, \cr
Q_4 =&\;
       \bar{u}^\alpha\,\gamma_\mu\,(1-\gamma_5)\,c_\beta \;
       \sum_q\;\bar{q}^\beta\,\gamma^\mu\,(1-\gamma_5)\,q_\alpha, \cr
Q_5 =& \;
       \bar{u}^\alpha\,\gamma_\mu\,(1-\gamma_5)\,c_\alpha\;
       \sum_q\;\bar{q}^\beta\,\gamma^\mu\,(1+\gamma_5)\,q_\beta, \cr
Q_6 =& \;
       \bar{u}^\alpha\,\gamma_\mu\,(1-\gamma_5)\,c_\beta \;
       \sum_q\;\bar{q}^\beta\,\gamma^\mu\,(1+\gamma_5)\,q_\alpha\;. \cr}}

The operator $Q_1^s$ ($Q_2^s$) in eq. \eHEFFCF ~is obtained from
$Q_1^d$ ($Q_2^d$) with
the substitution $(d \to s)$. In eqs. \eHEFFCA ~and
\eOPER ~$\alpha$ and $\beta$ are colour indices (that we will
omit in the following formulae) and in the ``penguin''
operators
$q$ ($\bar{q}$) is to be summed over all active flavours ($u$, $d$, $s$).

If we neglect mixing with the third generation ($U_{ub}=0$ and
$U_{us}\,U_{cs}^*=-U_{ud}\,U_{cd}^*= \sin \theta_C \cos \theta_C$) then
the three effective hamiltonians
\eqn\eSYM{
{H_{eff}^{\Delta C = \Delta S} \over \cos^2 \theta_C}
\;,\;\; {- H_{eff}^{\Delta S=0} \over \sqrt{2} \sin \theta_C \cos \theta_C}
\;,\;\; {H_{eff}^{\Delta C = - \Delta S} \over \sin^2 \theta_C}}
form a $U$-spin triplet.
Therefore, in the limit of exact flavour SU(3) symmetry a number
of relations between decay amplitudes should hold. We shall discuss
some of them in Section 4 and we will see that they are
often violated rather strongly.

We have evaluated the coefficients $C_i$ at the scale 1.5 GeV
using the two-loop anomalous dimension matrices recently calculated
by A. Buras and collaborators \rHEFNL, assuming
$\Lambda_4^{ \overline {MS}} = 300$  MeV. This value, which corresponds to
the best agreement between the experimental data and the theoretical results
on the exclusive decay channels of D mesons, is compatible with the
experimental determination from LEP measurements \rPDG.
The coefficients at next-to-leading order
are renormalization scheme dependent: we assume in the following the
values obtained using the "scheme independent prescription" of Buras
et al., namely $C_1 = -0.628$, $C_2 = 1.347$, $C_3 = 0.027$,
$C_4 = -0.057$, $C_5 = 0.015$, $C_6 = -0.070$.

In the factorized approximation the matrix elements of $H_{eff}$
are written in terms of matrix elements of {\it currents},
$(V_{q^{\prime}}^q)^{\mu}$ = $\bar{q}^{\prime}\,\gamma^{\mu}\,q$
and $(A_{q^{\prime}}^q)^{\mu}$ = $\bar{q}^{\prime}\,\gamma^{\mu}\gamma_5\,q$.

We recall the definitions of the decay constants for
pseudoscalar ($\pi$, $K$\dots) and vector ($\rho$, $K^*$\dots) mesons
\eqn\eFM{\eqalign{
  <P_i(p)|\,A^\mu\,|0> =& - i\;f_{P_i}\;p^\mu\ \ ,\cr
  <V_i(p,\lambda)|\,V^\mu\,|0> =& M_i\;f_{V_i}\;
          \epsilon^{*\mu}(\lambda) \ ,}}
and the usual definitions
\ref\rWSB{M. Wirbel, B. Stech and M. Bauer,
Zeits.f.Phys. C 29 (1985) 637.}
{}~for the matrix elements of the currents:
\eqn\eFOFA{\eqalign{
  <P_i|V^\mu|P_j> =& \;(p_i^\mu+p_j^\mu-{M_j^2-M_i^2 \over q^2}\,q^\mu)\,
  f_+(q^2) + \cr
  & + {M_j^2-M_i^2 \over q^2}\,q^\mu \,f_0(q^2),\cr
  <V_i|A^\mu|P_j> =& \;i\,(M_j+M_i)\,A_1(q^2)\,(\epsilon^{*\mu}-
  {\epsilon^* \cdot q \over q^2}\, q^\mu) - \cr
 & - i\,A_2(q^2) {\epsilon^* \cdot q \over M_j + M_i} \,
  (p_i^\mu+p_j^\mu-{M_j^2-M_i^2 \over q^2}\,q^\mu) +\cr
 & + i\,2\,M_i\,A_0(q^2)\,{\epsilon^* \cdot q \over q^2}\, q^\mu ,\cr
  <V_i|V^\mu|P_j> =&\; 2\,V(q^2)\,{{\varepsilon^\mu}_{\nu\rho\sigma}
  p_j^\nu\,p_i^\rho\,\epsilon^{*\sigma} \over M_j + M_i}.}}

To avoid the presence of a spurious singularity at $q^2$ = 0, one has to
require that
\eqn\eFPOLV{f_+ (0) = f_0 (0) \equiv v_{q q\prime},}
\eqn\eFPOLA{A_0 (0) = {M_i + M_j\over 2\,M_i} A_1 (0) +
  {M_i - M_j\over 2\,M_i} A_2 (0) \equiv a_{q q\prime}.}

The semileptonic decay rate for $D^0 \to K^- e^+ \nu$
\ref\rEXSEMI{A.L. Bean et al.
(Fermilab Tagged Photon Spectrometer Collaboration),
PANIC XII Conference, editors J.L. Matthews, T.W. Donnelly, E.H. Fahri
and L.S. Osborne, Nucl.Phys. A 527 (1991) 755c.}
indicates a value $v_{cs} \simeq 0.79$; assuming SU(3) symmetry for
the weak charges, we will set
\eqn\eQV{v_{cs} =v_{cd} =v_{cu} = 0.79}
Using the definition \eFPOLA ~and
the data on the semileptonic decay
 $D^0 \to K^{*-} e^+ \nu$  \ref \rEXSEMIV {E691 Collaboration:
J.C. Anjos et al., Phys.Rev.Lett. 65 (1990) 2630 \semi
E653 Collaboration: K.Kodama et al., Phys.Lett. B 274 (1992) 246
\semi E687 Collaboration:
 P.L. Frabetti et al., Phys.Lett. B 307 (1993) 262.}
we obtain $ a_{cs}
\simeq .54\pm.13$ (E653 Collaboration) or $ a_{cs} \simeq .71\pm.16$
(E691 Collaboration). Different lattice QCD calculations
\ref\rLATTICE{C.R. Allton et al., Nucl. Phys. B 416 (1994) 675 \semi
V. Lubicz et al. Phys.Lett. B 274 (1992) 415.} give similar
results: in average $ a_{cs} \simeq .74\pm0.15 $.
The limited statistics for the data on $D^0 \to \rho^- e^+ \nu$ decay
does not allow an analysis of the different form factors, but within
large errors the measured branching fraction is larger than theoretical
predictions based on quark models
\ref\rJAPP{E653 Collaboration: K.Kodama et al., Phys.Lett. B 316
(1993) 445.}.
Lattice results and results obtained in the framework
of QCD sum rules by using \eFPOLA ~
on SU(3) breaking are inconclusive because of large
errors on the $A_2$ form factor.

Since the data on $D$ meson decays show large SU(3) breaking effects
and since the axial charges are not protected by the so-called
Ademollo-Gatto theorem,
in our fit we allowed $a_{cs}$ and $a_{cd} = a_{cu}$ to vary between
.5 and 1 independently. The values chosen by the fit are $a_{cs} = 0.59$,
 consistent
with experimental data, and $a_{cd} = 1$ (in fact, an even better fit
would be obtained allowing larger values for $a_{cd}$).
We note that these values do not agree with the direct QCD sum rule
calculation of
$A_{0}(q^2)$ performed in \ref\rCFS{P. Colangelo, F. De Fazio and P.
Santorelli,
Preprint BARI-TH/94-174, DSF-T-94/12 to appear in Phys. Rev. D},
where the authors conclude that SU(3) breaking effects should be small
($a_{cs}/a_{cu}=1.10 \pm 0.05$) and the $q^2$
dependence of the form factors compatible with a polar dependence
dominated by the $0^-$ pole.

The $q^2$ dependence of the form factors is assumed to be dominated
by the nearest singularity. This entails for the $c$-quark decay terms
the usual simple-pole form factors
\eqn\eFFFF{\eqalign{
   f_+(q^2) =& {v_{cu(s,d)} \over 1\,-\,q^2/M^2_{D^*_{(s)}(1^-)}}\,,\cr
   f_0(q^2) =& {v_{cu(s,d)} \over 1\,-\,q^2/M^2_{D^*_{(s)}(0^+)}}\,,}}
and analogous expressions, with the mass of the lightest
particle with appropriate quantum numbers, for the other form factors.

In the W-exchange and annihilation terms, however, the large and time-like
$q^2$ values needed, together with the suggested existence of resonances
with masses near to the $D$-meson mass, make a prediction based on the
lightest mass singularity unjustified. These terms depend on the matrix
elements of
current divergences between the vacuum and two-meson states.
We write them, with the help of the equations of motion, in the
way indicated in the following examples:
\eqn\eDIVER{\eqalign{
       <K^-\pi^+|\partial^\mu(V_s^d)_\mu|0> =&
      \;i\,(m_s-m_d)\,<K^-\pi^+|\,\bar{s}d\,|0> \equiv \cr
 \equiv &  \;i\,(m_s-m_d)\,{M_D^2 \over f_D}\,W_{PP}, \cr
       <K^-\rho^+|\partial^\mu(A_s^d)_\mu|0> =&
      \;i\, (m_s+m_d)\,<K^- \rho^+|\,\bar{s}\gamma_5 d\,|0> \equiv \cr
 \equiv &
    -\, (m_s+m_d)\,{2\,M_\rho \over f_D}\,\epsilon^*\cdot p_K\,W_{PV}.\cr}}
We assume SU(3) symmetry for the matrix elements of scalar and
pseudoscalar densities, and express all of them in terms of $W_{PP}$,
$W_{PV}$. In our approach the $W_i$'s are
free parameters of the fit. Their magnitude turns out to be
considerably larger than what one would obtain assuming
form factors dominated by the pole
of the lightest scalar or pseudoscalar meson, i.e. $K_0^*(1430)$ or
$K(497)$.

We note that to obtain the amplitudes for Cabibbo first-forbidden decays
one has to evaluate matrix elements like $<\eta|\bar d \gamma^\mu
\gamma_5 d|0>$ or, for penguin operators, $ <\eta|\bar d \gamma_5 d|0>$.
To get the correct result it is necessary to take into account the
anomaly of the singlet axial current; we have followed the method
discussed in \ref\rETAETAP{D.I. D'yakonov and M.I. Eides, Sov.Phys.JETP
54 (1981) 232 \semi G. Veneziano, Nucl. Phys. B 159 (1979) 213 \semi
I. Halperin, Phys. Rev. D 50 (1994) 4602.}.
In this scheme the $\eta-\eta^{\prime}$ mixing angle,
$\theta_{\eta\eta^{\prime}}$,
results to be equal to $-10^{\circ}$. Remarkably,
this value which is consistent with
the Gell-Mann Okubo mass formula, is also perfectly compatible with the
experimental value of $\Gamma(\eta \rightarrow \gamma \gamma)$ obtained
by two-photon production experiments \ref\rETAGAM{Williams et al., Phys.Rev.
D 38 (1988) 1365}. Therefore the $\eta-\eta^{\prime}$ mixing angle
is not a parameter of the fit, as it was in our previous analyses.

If the final $K$ meson in Cabibbo allowed decays is neutral, it  has
been observed as a short-lived neutral $K$, $K_S$. There is therefore an
interference between Cabibbo allowed ($D\rightarrow \KB + X$) and doubly
suppressed ($D\rightarrow K^0 + X$) decay amplitudes. We have fitted the
experimental data $\Gamma_{exp}$ \rPDG~ using the definition
\eqn\eKL{
\Gamma (D \rightarrow K_S +X) \equiv
{1 \over 2}\;\Gamma_{exp} (D \rightarrow \KB +X)\,,}
and included FSI modifications also for the doubly Cabibbo suppressed
part of the amplitude.
The correction due to this effect is not negligible
and it helps in
obtaining a better fit to the experimental data.

We write now a few examples of  amplitudes for Cabibbo allowed
and first forbidden decays
in the factorized approximation:
\eqn\eAKMPP{\eqalign{
{\cal A}_{\rm w}(D^0 \to K^- \pi^+) &= \cr
     = - {G_F \over \sqrt 2}\,U_{ud}\,U_{cs}^*
  &\phantom{=\,}
   \bigl[(C_2+\xi\,C_1)\;<K^-|(V_s^c)_\mu|D^0>\;<\pi^+|(A_u^d)^\mu|0> +\cr
    &\phantom{=\,}
 + (C_1+\xi\,C_2)\;<K^-\pi^+|(V_s^d)_\mu|0>\;<0|(A_u^c)^\mu|D^0>\, \bigr] =\cr
  =  - {G_F \over \sqrt 2}\,U_{ud}\,U_{cs}^*
 &\phantom{=\,}
   \bigl[(C_2+\xi\,C_1)\;f_\pi\;<K^-|\partial^\mu(V_s^c)_\mu|D^0> +\cr
 &\phantom{=\,}
 + (C_1+\xi\,C_2)\;f_D\;<K^-\pi^+|\partial^\mu(V_s^d)_\mu|0>\, \bigr].\cr}}
In \eAKMPP  ~the two terms correspond to $c$-quark
decay and W-exchange, respectively .
\eqn\eAKSRHP{\eqalign{
{\cal A}_{\rm w}(D^+ \to K_S \rho^+) &= \cr
      = -\,  {G_F \over 2}\,U_{ud}\,U_{cs}^*
  &\phantom{=\,}
   \bigl[(C_2+\xi\,C_1)\;<\KB|(V_s^c)_\mu |D^+>\;
  <\rho^+|(V_u^d)^\mu|0> +\cr
    &\phantom{=\,}
  + (C_1+\xi\,C_2)\;<\KB|(A_s^d)_\mu |0>\;
  <\rho^+|(A_u^c)^\mu|D^+>\, \bigr] +\cr
     +\,  {G_F \over 2}\,U_{us}\,U_{cd}^*
  &\phantom{=\,}
   \bigl[(C_2+\xi\,C_1)\;<0|(A_d^c)_\mu |D^+>\;
  <K^0\rho^+|(A_u^s)^\mu|0> +\cr
    &\phantom{=\,}
  + (C_1+\xi\,C_2)\;<K^0|(A_d^s)_\mu |0>\;
  <\rho^+|(A_u^c)^\mu|D^+>\, \bigr] =\cr
 =   -\, {G_F \over 2}\,U_{ud}\,U_{cs}^*
  &\phantom{=\,}
   \bigl[(C_2+\xi\,C_1)\;f_\rho\,M_\rho\,\epsilon^{*\mu}(\rho^+)\;
  <\KB|(V_s^c)_\mu |D^+> + \cr
  &\phantom{=\,}
  + (C_1+\xi\,C_2)\;f_K\;<\rho^+|\partial^\mu(A_u^c)_\mu|D^+>\, \bigr] +\cr
     +\,  {G_F \over 2}\,U_{us}\,U_{cd}^*
  &\phantom{=\,}
   \bigl[(C_2+\xi\,C_1)\;f_D\;
  <K^0\rho^+|\partial^\mu(A_u^s)_\mu|0>\,+ \cr
  &\phantom{=\,}
  + (C_1+\xi\,C_2)\;f_K\;<\rho^+|\partial^\mu(A_u^c)_\mu|D^+>\,\bigr].\cr}}
In \eAKSRHP ~we have given an example of an amplitude with $\KB$ and $K^0$
interfering contributions, that also contains an annihilation term.
\eqn\eAPOPP{\eqalign{
{\cal A}_{\rm w}(D^+ \to \pi^0 \pi^+) &= \cr
      = - {G_F \over \sqrt 2}\,U_{ud}\,U_{cd}^*
  &\phantom{=\,}
   \bigl[(C_2+\xi\,C_1)\;<\pi^0|(V_d^c)_\mu|D^+>\;<\pi^+|(A_u^d)^\mu|0> +\cr
    &\phantom{=\,}
+ (C_1+\xi\,C_2)\;<\pi^+|(V_u^c)_\mu|D^+>\;<\pi^0|(A_d^d)^\mu|0>\, \bigr] =\cr
 =  - {G_F \over \sqrt 2}\,U_{ud}\,U_{cd}^*
  &\phantom{=\,}
   \bigl[(C_2+\xi\,C_1)\;f_\pi\;<\pi^0|\partial^\mu(V_d^c)_\mu|D^+> +\cr
  &\phantom{=\,}
 - (C_1+\xi\,C_2)\;{f_{\pi} \over \sqrt{2}}
 \;<\pi^+|\partial^\mu(V_u^c)_\mu|D^+>\, \bigr]\;= \cr
=  + {G_F \over 2}\,U_{ud}\,U_{cd}^*
&\phantom{=\,} (C_1 + C_2)\;(1+\xi)\;f_{\pi}
\;<\pi^+|\partial^\mu(V_u^c)_\mu|D^+>\;.\cr}}
\eqn\eAPMPP{\eqalign{
{\cal A}_{\rm w}(D^0 \to \pi^- \pi^+) &= \cr
     = - {G_F \over \sqrt 2}\,U_{ud}\,U_{cd}^*
  &\phantom{=\,}
     (C_2+\xi\,C_1)\;<\pi^-|(V_d^c)_\mu|D^0>\;<\pi^+|(A_u^d)^\mu|0> +\cr
      + {G_F \over \sqrt 2}\,U_{ub}\,U_{cb}^*
    &\phantom{=\,}
   \bigl[(C_4+\xi\,C_3)\;<\pi^-|(V_d^c)_\mu|D^0>\;<\pi^+|(A_u^d)^\mu|0> +\cr
    &\phantom{=\,}
   - 2\,(C_6+\xi\,C_5)\;<\pi^- \pi^+|\bar {u} u|0>\;
  <0|\bar u \gamma_5 c|D^0> +\cr
    &\phantom{=\,}
   + 2\,(C_6+\xi\,C_5)\;<\pi^-|\bar d  c|D^0>\;
  <\pi^+|\bar u \gamma_5 d|0> \, \bigr] =\cr
  =  - {G_F \over \sqrt 2}\,U_{ud}\,U_{cd}^*
 &\phantom{=\,}
     (C_2+\xi\,C_1)\;f_\pi\;<\pi^-|\partial^\mu(V_d^c)_\mu|D^0>\; +\cr
      + {G_F \over \sqrt 2}\,U_{ub}\,U_{cb}^*
    &\phantom{=\,}
   \bigl[(C_4+\xi\,C_3)\;f_\pi\;<\pi^-|\partial^\mu(V_d^c)_\mu|D^0>\; +\cr
    &\phantom{=\,}
   + 2\,i\,(C_6+\xi\,C_5)\;<\pi^- \pi^+|\bar {u} u|0>\;
  {{f_D\;M_D^2} \over {m_u + m_c}} +\cr
    &\phantom{=\,}
   - 2\,i\,(C_6+\xi\,C_5)\;<\pi^-|\bar d  c|D^0>\;
  {{f_{\pi}\;M_{\pi}^2} \over {m_u + m_d}} \, \bigr] .\cr}}
In \eAPOPP ~and \eAPMPP ~we give examples of Cabibbo forbidden decay
amplitudes, to the second of which penguin operators contribute.

The parameter $\xi$ appearing in the above equations should be equal
to ${1 \over 3}$ in QCD. Since however other colour suppressed,
nonfactorizable contributions have been neglected we will consider it as
a free parameter
to be fitted to the
experimental data, following \rBSW .
The result of the fit favours a value $\xi \simeq 0$.

\newsec{Final state interaction effects}
We make the assumption that FSI are dominated by resonant contributions,
and we neglect the phase-shifts in exotic channels.
In the mass region of pseudoscalar charmed particles there is evidence,
albeit not very strong \rPDG , for
a $J^P=0^-$ $K(1830)$ (with $\Gamma = 250$ MeV and
an observed decay to $K \phi$
\ref\rARMST{T. Armstrong et al., Nucl. Phys. B 221 91983) 1.})
and a $J^P=0^-$ $\pi(1770)$ with $\Gamma = 310$ MeV
\ref\rBELL{G. Bellini et al., Phys.Rev.Lett. 48 (1982) 1697.}.
The coupling of an octet of $0^-$ $\PT$ resonance to $0^- + 1^-$ ($PV$)
channels is determined from charge conjugation and SU(3) symmetry to be:
\eqn\eGPPV{h\; f_{abc}\,(\partial_\mu\,{\PT}_a)\,V^\mu_b\,
          P_c\,.}
Equation \eGPPV ~implies:
\eqn\eWIDTHS{{\Gamma(\widetilde K \to PV) \over
    \Gamma(\widetilde \pi \to PV)} = 0.8 \simeq
    {\Gamma_{exp}(\widetilde K) \over \Gamma_{exp}(\widetilde \pi)},}
consistent with the assumption that the $\PT$
resonances decay predominantly into the lowest lying $P$ and $V$ mesons,
which we shall make for simplicity.

For Cabibbo allowed and doubly forbidden $D^0$ decays
and for Cabibbo first and doubly forbidden $D^+$ decays,
the FSI effect modifies the amplitudes in the following way:
\eqn\eRESCA{{\cal A} (D \to V_h\,P_k) = {\cal A}_{\rm w}
  (D \to V_h\,P_k) + c_{hk}[\exp(i\delta_8)-1]
   \sum_{h\prime k\prime} c_{h\prime k\prime}\,{\cal A}_{\rm w}
  (D \to V_{h\prime}\,P_{k\prime})\,,}
In \eRESCA ~ $c_{hk}$ are the normalized ($\sum c_{hk}^2 = 1$)
couplings $\widetilde P P V$ and
\eqn\ePHSH{\sin \delta_8 \, \exp (i \delta_8) =
   {\Gamma ({\PT} ) \over 2\,(M_{\PT} - M_D) - i\,\Gamma ({\PT} )},}
where $\PT$ is the resonance appropriate to the decay channel considered
($\widetilde \pi$ or $\widetilde K$).
Equation \ePHSH ~determines $\delta_8$ up to a $180^{\circ}$ ambiguity.
The choice can be made according to the number of resonances and bound
states
\ref\rLEVINS{N. Levinson, Kgl. Danske Vid. Selskab, Mat.-fys. Medd.,
25 (1949).}
that are present in the channel at lower energies, each one of them increasing
the phase-shift by $\pi$. In this case the $\PT$ resonance
may be assumed as the third resonance in the
channel, with  $\delta_8$ close to ${1 \over 2} \pi$ (in our recent preprint
a different choice was made for the $PV$ channels).

For Cabibbo forbidden $D^0 \to PV$ decays one expects
$\widetilde \eta$ and $\widetilde{\eta^{\prime}}$ resonances to
take also part to rescattering. Due to charge conjugation invariance,
the singlet components have vanishing coupling and
the combined effect of the two expected isoscalar resonances may be
described by a phase attached to the isosinglet octet part of
the decay amplitudes.
This phase is the only added parameter for these channels to be fitted.
The fitted value is $\sim 243^{\circ}$, corresponding to two
resonances with masses one below and the other above the $D^0$ mass.

Coming now to FSI effects for parity violating
$D \to PP$ decays, we note that some
evidence exists for
a $J^P=0^+$ resonance $K^*_0$ (with mass
1945$\pm$10$\pm$20 MeV, width 201$\pm$34$\pm$79 MeV and 52$\pm$14\%
branching ratio in $K \pi$
\ref\rASTON{Aston et al., Nucl.Phys. B 296 (1988) 493.}).
 No $a_0$ isovector resonance has been
observed up to now in the interesting mass region.
In \rBUCCA ~we assumed its existence and we estimated its mass from
the equispacing formula
\eqn\eEQUISP{M_{a_0}^2 = M^2_{K^*_0}-M^2_K+M^2_\pi .}
In the fit we allowed the mass and the width of $K^*_0$ to vary
within the experimental bounds. The best fit values are 1928
MeV and 300 MeV, respectively. From Eq. \eEQUISP ~we get $M_{a_0}=$
1869 MeV.

The SU(3) and $C$ invariant coupling of a scalar $S$ resonance
to two pseudoscalar mesons is:
\eqn\eGPPP{\eqalign{
        \sqrt{3 \over 2}\,g_{888}\,d_{abc}\,P_a\,P_b\,S_c +&
          g_{818}\,(\,P_a\,P_0+P_0\,P_a\,)\,S_a\,+ \cr
    + g_{881}\,P_a\,P_a\,S_0 +& g_{111}\,P_0\,P_0\,S_0.}}
In  \eGPPP ~$a,b,c=1\ldots 8$ are SU(3) indices, $P_0$ and $S_0$ are
SU(3) singlets.

Assuming that the $S$ resonances decay dominantly to a pair of mesons
belonging to the lowest mass nonet, one obtains from \eGPPP ~the branching
ratio for $(K^*_0 \to K \pi)$ as a function of
the ratio of coupling
constants $r = g_{818}\, /\, g_{888}$. The further assumption of nonet
symmetry would imply $r$ = 1. The experimental data allow two possible
values for $r$, one positive (and consistent with 1) and another negative
and close to $- 1$.
The best fit value of $r$ is $-0.84$, corresponding to a
branching ratio for the decay of the $K^{*}_{0}$ resonance in $K \pi$ of
about $64\%$.

The description of rescattering effects for Cabibbo forbidden $D^0$ decays
is complicated by the presence of yet unobserved $f_0$ and $f_0^{\prime}$
isoscalar resonances, which should be singlet-octet mixtures.
Denoting by $|f_0 \rangle$ the lower mass state, we define
\eqn\eSCALAR{\eqalign{
  |f_0\rangle =& \phantom{ -} \sin \phi \; |f_8\rangle + \cos \phi \;
|f_1\rangle \cr
   |f_0^{\prime}\rangle =& -\cos \phi \; |f_8\rangle + \sin \phi \;
|f_1\rangle.}}
The results will also depend on the parameters
$a = g_{881}\, /\, g_{888}$ and $c = g_{111}\, /\, g_{888}$, see \eGPPP .

To reduce the number of new parameters, we assume that these scalar resonances
behave similarly to the tensor mesons
$f_2\;(1270)$ and $f_2^{\prime}\;(1525)$:
the $f_2^{\prime}$ is very weakly coupled to $\pi \pi$,
and the
$f_2$ has in turn a small coupling to $K \overline{K}$.
In order to forbid the
$f_0^{\prime} \to \pi \pi$ decay, we required $a$ and the mixing angle $\phi$
to be related by
$$a = {1 \over{ \sqrt 2 \ \tan \phi} \;}.$$
The value $\tan \phi = \sqrt 2$ would then imply
a vanishing branching ratio for the decay
$f_0 \to K \overline{K}$. The
best fit value is $\tan \phi = 1.14$, not very far from $1.41$.

The fit has therefore two new free parameters,
$\Delta^2 = M_{f_0^{\prime}}^2 - M_{f_0}^2$ and  $\phi$. For any
pair ($\Delta^2, \phi$)
there are two possible values for $c$, that are solutions of
a quadratic condition coming from the requirement of unitarity of the
rescattering transformation.

\newsec{Comparison with experimental data on Branching Ratios}
Starting from the weak amplitudes $\cal A_{\rm w}$ defined in Section 2
and modifying them with FSI effects as explained in the previous Section,
we evaluated the rates for all Cabibbo allowed two-body decays
and for Cabibbo forbidden $D^+$ and $D_s^+$ decays as
functions of the parameters of the fit. These are $\xi$,
the parameters $W_{PP}$ and $W_{PV}$ of the annihilation contributions,
the axial charges $a_{cu}=a_{cd}$ and $a_{cs}$,
the mass and width of the scalar $K^{*}_{0}$ resonance. We use the values
($m_u,\,m_d,\,m_s,\,m_c$) = (4.5, 7.4, 150, 1500 MeV) for the quark
masses. The decay constants are
($f_{\pi},\,f_K,\,\,f_{\rho}=f_{K^*},\,f_{\omega},\,f_{\phi}$) =
(133, 160, 216, 156, 233~MeV). The pole masses in the form factors \eFFFF ~
corresponding to yet undetected charmed particles have been taken to be
$M_{D^*(0^+)} = 2470 \,{\rm MeV}$ and $M_{D^*_s(0^+)} = 2600 \,{\rm MeV}$.
Other parameters relevant for decay to final states containing $\eta
(\eta^{\prime})$ have been fixed following \rETAETAP .
The results are presented in Tables 1, 2, 3.

The best-fit results are reported in column three of the Tables.
The total $\chi^2$ is 80.6. The 25 data points
for Cabibbo allowed decays contribute 61.8 and the 12 data points
for Cabibbo first forbidden decays 18.8. The best fit parameters are
$\xi=-0.027$, $r = -0.84$,
$W_{PP}=-0.29 $, $W_{PV}=+0.29$, $a_{cu}=a_{cd}=1.0$, $a_{cs}=0.59$,
$M_{K^*_0}=1928\;{\rm MeV}$ and $\Gamma_{K^*_0}=300\;{\rm MeV}$.
A separate fit to Cabibbo allowed data alone
gives quite similar values for the parameters. In the Tables we have also
reported the theoretical predictions for the decays to final states
containing $K_L$, in order to show the importance of interference effects
with doubly Cabibbo-suppressed amplitudes.

We note that the relatively large SU(3) violation present in the data
for exotic $D^+$ decays \rPDG
\eqn\eDPIUSU{R_+= \left|{U_{cs} \over U_{cd}}\right|^2
{\Gamma(D^+ \to \pi^+ \pi^0) \over \Gamma(D^+ \to K_S \pi^+)} \simeq
3.55,}
(instead of $R_+ =1$) is well reproduced by the fitted data, as it also
happened in \rBUCCB . This point has been discussed in detail in
\ref\rMARBEL{A. Pugliese and P. Santorelli, preprint INFN-NA-IV-93-26,
in "Third Workshop on the Tau-Charm Factory",
J. Kirkby and R. Kirkby eds., Editions Fronti\`eres.} ~and more
recently in
\ref\rCHAU{L.-L. Chau and H.-Y. Cheng, preprint ITP-SB-93-49/UCD-93-31.}.
The reason is that several SU(3) breaking effects, each one rather
small, contribute coherently to enhance $R_+$.

We have also evaluated, not including them in the fit, two recently measured
doubly Cabibbo forbidden branching ratios. The experimental data are
\ref\rCLEODCS{CLEO Collaboration: D. Cinabro et al. Phys. Rev. Lett.
72 (1994) 406.},
\ref\rANJOS{J.C. Anjos et al., Phys.Rev.Lett. 69 (1992) 2892.}
\eqn\eDDKPI{R_0^{K \pi} = {B.R.(D^0 \to \pi^- K^+) \over B.R.(D^0 \to
\pi^+ K^-)} =
(0.77 \pm 0.25 \pm 0.25)\;10^{-2}}
\eqn\eDDPHI{R_+^{\prime} = {B.R.(D^+ \to \phi K^+) \over
B.R.(D^+ \to \phi \pi^+)} =
(5.8^{\displaystyle +3.2}_{\displaystyle -2.6} \pm 0.7)\;10^{-2}\;,}
and the theoretical predictions using the best-fit parameters
are $R_0^{K \pi} = 0.89\;10^{-2}$and \break
$R_+^{\prime} = 0.76\;10^{-2}$.

In the SU(3) limit the U-spin properties of
the hamiltonian should give
$R_0^{K \pi} = \tan^4 \theta_C = 0.26\;10^{-2}$,  also
discussed in \rCHAU . Our model predicts $R_0^{K \pi}$ to be 3.4 times
larger than this value. This is due mainly to the W-exchange
contributions, that should vanish in the symmetric limit and
have opposite signs in the two amplitudes, and also to
rescattering effects.

On the other hand the theoretical value for $R_+^{\prime}$
is much smaller than the experimental datum, that however differs
from zero by only 2.5 standard deviations. We note that
in a factorized model the decay $D^+ \to \phi K^+$
may only proceed through annihilation
and rescattering. It is therefore difficult to reproduce the
present very large value for $R_+^{\prime}$.

We consider now $D^0$ decay processes, and in particular
Cabibbo first-forbidden decays.
The $D^0$ meson
is a $U$-spin singlet and it should only decay to $U$-spin triplet states,
if flavour SU(3) is a good symmetry.
Therefore in that limit several relations among decay
amplitudes hold. For the parity violating $D^0 \to PP$ decays they are:
\eqn\eUSPINA{{\cal A}(D^0 \to K^0 \KB) = \;0\;,}
\eqn\eUSPINB{\eqalign{
{\cal A}(D^0 \to K^+ K^-) =& - {\cal A}(D^0 \to \pi^+ \pi^-) = \cr
= {{\cal A}(D^0 \to K^+ \pi^-) \over \tan \theta_C}
=& - \tan \theta_C {\cal A}(D^0 \to K^- \pi^+)\;}}
(the last equality corresponds to the symmetry prediction discussed after
\eDDKPI ),
\eqn\eUSPINC{{1 \over \sqrt{3}}{\cal A}(D^0 \to K_S \pi^0) =
\cos \theta_{\eta \eta^{\prime}} {\cal A}(D^0 \to K_S \eta) +
\sin \theta_{\eta \eta^{\prime}} {\cal A}(D^0 \to K_S \eta^{\prime}) \;,}
\eqn\eUSPIND{
-{1 \over \sqrt{3}}{\cal A}(D^0 \to \pi^0 \pi^0) = \;
\cos \theta_{\eta \eta^{\prime}} {\cal A}(D^0 \to \pi^0 \eta) +
\sin \theta_{\eta \eta^{\prime}} {\cal A}(D^0 \to \pi^0 \eta^{\prime})\;,}
\eqn\eUSPINE{
- {\cal A}(D^0 \to \pi^0 \pi^0) = \;
(1 - \tan^2 \theta_{\eta \eta^{\prime}}) {\cal A}(D^0 \to \eta \eta) +
\;2\;\tan \theta_{\eta \eta^{\prime}} {\cal A}(D^0 \to \eta \eta^{\prime})\;,}
\eqn\eUSPINF{\eqalign{
{\cal A}(D^0 \to \eta \eta^{\prime}) =& \tan \theta_{\eta \eta^{\prime}}\;
{\cal A}(D^0 \to \eta \eta) +  \cr
+&\,\sqrt{3}\;\bigl(\sin \theta_{\eta \eta^{\prime}}
{\cal A}(D^0 \to \pi^0 \eta) - \cos \theta_{\eta \eta^{\prime}}
{\cal A}(D^0 \to \pi^0 \eta^{\prime})\bigr)\;.}}
Comparing the above formulae to the experimental data \rPDG , \rCLEODP ,
\rCLEODCS , we note that \eUSPINA ~is definitely not true, the moduli
of the amplitudes in \eUSPINB ~are in the ratios (1:0.59:1.15:0.67)
instead of being equal, relation \eUSPINC ~is compatible with the data,
but only with large phases, and finally no data exist for \eUSPIND ,
\eUSPINE , \eUSPINF .

In the factorization scheme the amplitude ${\cal A}(D^0 \to K^0 \KB)$
vanishes, and we may only obtain a nonzero rate in that channel through
rescattering from the other decay channels. Analogously, the small SU(3)
breaking effects ($f_{\pi} \neq f_K$, $M_D^2\,-\,M_{\pi}^2\,>\,
M_D^2\,-\,M_K^2$,$\ldots$) are not enough to reproduce the large ratio
of $K^+ K^-$ to $\pi^+ \pi^-$ rates. In Table 4 we show the results of
our best fit to the Cabibbo first-forbidden decay rates.
As explained in Section 3,
to the parameters previously determined we added two more
free parameters, $\Delta$ and $\phi$, and chose the value of another one, $c$,
between the two solutions of a quadratic consistency condition.
The fit we obtain is good
($\chi^2 = 1.7$ with 4 data points). The best fit results for the
parameters are
$\Delta = \sqrt{M_{f_0^{\prime}}^2 - M_{f_0}^2} = 1205\,{\rm MeV}$,
$\phi = 48.7^{\circ}$ and $c \simeq -2.69$. The
corresponding masses and widths of the scalar and isoscalar
resonances are $M_{f_0} = 1778\,{\rm MeV}$, \break
$\Gamma_{f_0} = 361\,{\rm MeV}$,
$M_{f_0^{\prime}} = 2148\,{\rm MeV}$
and $\Gamma_{f_0^{\prime}} = 389\,{\rm MeV}$.

For parity conserving decays, the relations obtained from flavour
symmetry are less restrictive in view of the few existing data. For
instance, the relation corresponding to \eUSPINA , namely
\eqn\eUSPINVEC{{\cal A}(D^0 \to K^0 \overline {K}^{*0})
+ {\cal A}(D^0 \to \KB K^{*0}) = \;0\;,}
is valid in the factorization approximation and rescattering does
not spoil its validity in our model. Only upper limits exist experimentally.
\hfill\break
Similarly, other SU(3) predictions are
\eqn\eUSPINFIN{\eqalign{
{\cal A}(D^0 \to K^+ K^{*-}) =& \,-\,{\cal A}(D^0 \to \pi^+ \rho^-) \cr
{\cal A}(D^0 \to K^- K^{*+}) =& \,-\,{\cal A}(D^0 \to \pi^- \rho^+)\;.}}
The W-exchange terms strongly violate these relations
and our predictions are therefore at variance with them. Data for $\pi \rho$
final states are unfortunately still missing.

The experimental data and the fitted values for
parity conserving Cabibbo first-forbidden $D^0 \to PV$ decays are also
reported in Table 4.
The only parameter added to those determined in the previous fits
is the phase-shift of the isosinglet octet part of the decay amplitudes.
This parameter turns out to be $243^{\circ}$ and the quality of
the fit is reasonably good ($\chi^2 = 7.7$ for 4 data points).

\newsec{CP violation}
It is well known that CP violating effects show up in a decay process
only if the decay amplitude is the sum of two different
parts, whose phases are made of a weak (CKM) and a strong
(final state interaction) contribution.
The weak contributions to the phases change sign when going to the
CP-conjugate process, while the strong ones do not.
Let us denote a generic decay amplitude of this type by
\eqn\eAMPL{{\cal A}= A\; e^{i\delta_1} + B\; e^{i\delta_2}}
and the corresponding
CP conjugate amplitude by
\eqn\eAMPB{{\bar{\cal A}}= A^* e^{i\delta_1} + B^* e^{i\delta_2}\,.}
The CP violating asymmetry in the decay rates will be therefore
\eqn\eCPV{a_{CP} \equiv
{|{\cal A}|^2-|{\bar{\cal A}}|^2 \over |{\cal A}|^2+|{\bar{\cal A}}|^2}
= {2\;\Im(A B^*) \, \sin(\delta_2-\delta_1) \over
|A|^2+|B|^2+2\;\Re(AB^*)\,\cos(\delta_2-\delta_1)}.}
Both factors in the numerator of eq. \eCPV ~should be nonvanishing to have
a nonzero effect. Moreover, to have a sizeable asymmetry the moduli of the
two amplitudes $A$ and $B$ should not differ too much.

In Cabibbo first-forbidden $D$ decays, the penguin terms in the effective
hamiltonian \eHEFFCF ~provide the different phases of the weak amplitudes
$A$ and $B$. Having obtained a reasonable description of the decay
processes, including
a model for their strong phases, we may envisage the more ambitious goal to
derive CP violating asymmetries using our model for the phase-shifts.
The asymmetries resulting are around $10^{-3}$, somewhat larger than
previously expected. We stress however that the actual numbers may vary
appreciably for parameter variations that still give reasonable fits to
the decay rates.

For $D^+$ (and $D_s^+$) decays the total charge allows to directly
measure the rates to be combined in the asymmetry \eCPV . In the neutral
$D$ decays, however, the need of tagging the decaying particle to tell
its charm and the possibility of $D - \bar D$ mixing make \eCPV ~, as it
stands, not directly measurable.
Let us consider the particular case of an
$e^+ e^-$ factory at the $\psi^{\prime \prime}$, assume to tag
$D^0$ by semileptonic decays and define
\eqn\eFACTORY{a_{CP}^{eff.} =
{|{\cal A}(f_i,e^-)|^2-|{{\cal A}(\overline{f_i},e^+)}|^2
\over |{\cal A}(f_i,e^-)|^2+|{{\cal A}(\overline{f_i},e^+)}|^2}.}
Limiting our considerations to the simplest case, $D$ decays to a
CP self-conjugate final state ($f_i = \pm \overline{f_i}$), and
considering the time-integrated asymmetry, one obtains
\eqn\eTIMEINT{a_{CP}^{eff.} =
{{\strut\Bigl( 1 - \left|{\displaystyle p \over \displaystyle q}\right|^2
\Bigr)\,
\Bigl(1 + \left|{\displaystyle q \overline {\cal A} \over
\displaystyle p {\cal A}}\right|^2\Bigr) +
{\displaystyle {1-y^2} \over \displaystyle {1+x^2}}\;
\Bigl(1 + \left|{\displaystyle p \over \displaystyle q}\right|^2\Bigr)\,
\Bigl(1 - \left|{\displaystyle q \overline {\cal A} \over
\displaystyle p {\cal A}}\right|^2 \Bigr)}   \over
{\strut \Bigl(1 + \left|{\displaystyle p \over \displaystyle q}\right|^2
\Bigr)\,
\Bigl(1 + \left|{\displaystyle q \overline {\cal A} \over
\displaystyle p {\cal A}}\right|^2\Bigr) +
{\displaystyle {1-y^2} \over \displaystyle {1+x^2}}\;
\Bigl(1 - \left|{\displaystyle p \over \displaystyle q}\right|^2\Bigr)\,
\Bigl(1 - \left|{\displaystyle q \overline {\cal A} \over
\displaystyle p {\cal A}}\right|^2 \Bigr)}}.}
In \eTIMEINT ~${\cal A} = {\cal A}(D^0 \to f_i )$ is the decay amplitude,
the mass matrix eigenstates are defined as
$$|D_{S,L}\rangle \propto |D^0\rangle \pm {q \over p}\;|\overline
{D}^0\rangle$$
and the mixing parameters are
$$x={{2\;(M_S - M_L)} \over  {\Gamma_S + \Gamma_L}},\ \ \ \
y={{\Gamma_S - \Gamma_L} \over {\Gamma_S + \Gamma_L}}.$$

The mixing for charmed mesons is experimentally known to be small
(${|x| < 0.083}$, ${y < 0.085}$) \rPDG ~and the theoretical calculations of
the short-distance contributions give {\it very} small predictions. A
reliable calculation of long-distance terms is problematic,
even more so for CP violation in the mass matrix and the
ratio ${p/ q}$.
We did not consider at all the time-dependence in the asymmetries,
since these depend on the phase of ${p / q}$. We note anyhow
that the smallness of $x$ and $y$ prevents the development in time of
appreciable asymmetries even if the phases would allow this.
We expect moreover
that the modulus $|{p / q}|$ will differ from one by a small
amount, $O(10^{-3})$.
Therefore we can expand \eTIMEINT ~in the small quantities
$$a_{CP}={1 \over 2}\,\Bigl(1 - \Bigl|{\overline {\cal A} \over
{\cal A}}\Bigr|^2 \Bigr) \ \ \ \ \ {\rm and}\ \ \ \ \
\Re\epsilon = \;{1 \over 4}\Bigl(\Bigl|{p \over q}\Bigr|^2\,-\,1\Bigr)\;,$$
with the result
\eqn\eTIMEAPPR{a_{CP}^{eff.} \simeq \;a_{CP}\;(1-\delta)\,-\,2\;\delta\;
\Re\epsilon\;+\;\ldots}
In \eTIMEAPPR ~we have used the definition
$\delta=\;1\;-\;(1-y^2)\;/\; (1+ x^2)$. Experimentally \break
$\delta\,<\,0.014$.
Therefore, if $a_{CP}$ is not much smaller than $\Re\epsilon$, the
measured asymmetry will be ${a_{CP}^{eff.} \simeq a_{CP}}$.

We report in Table 5 the values of $a_{CP}$ that we obtain in our model
for several decay processes. We chose to give only the results
that correspond to a good fit for the branching ratios, even if the
predictions for some other decay channels are also large.
For parity conserving $D^0$ decays the final states are not CP eigenstates,
and the corresponding formula for $a_{CP}^{eff.}$ is more complicated;
however, we note that for amplitudes of not too different absolute values,
as it happens in Cabibbo first-forbidden decays, and given
the smallness of the mixing parameters, the result is again
${a_{CP}^{eff.} \simeq a_{CP}}$. If the final state
contains a neutral kaon, normally identified by its decay to
$\pi^+ \pi^-$, one has to disentangle the CP violating effects in $D$
and $K$ decays. How to do this for the $D^+$ decays has been discussed
in \rBUCCB . We note that these channels are not very promising
candidates to look for CP violating effects in $D$ decays.

We evaluated the central value for $a_{CP}$ choosing for the Maiani --
Wolfenstein parameters ($\rho$,$\eta$) the values ($0.2$, $0.3$), following
a recent analysis of CKM parameters
\ref\rCIUCHINI{ M. Ciuchini, E. Franco, G. Martinelli and L. Reina,
Roma preprint n. 1024-1994.} ,
and $U_{cb}$ = $0.040$. We varied ($\rho$,$\eta$) in the one-sigma region
obtained in \rCIUCHINI ~for $f_B = 200 \pm 40\;{\rm MeV}$.
The error given in Table 5 reflects
{\it only} this uncertainty, which is already quite large.

\newsec{Conclusion}
We have presented a generally successfull description of the complex of
two-body nonleptonic decays of charmed mesons. We fitted 45 experimental
branching ratios with eleven free parameters,
that assume values close to
the expected ones at the minimum of the $\chi^2$ = $90$.

We note that the large SU(3) breaking effects shown by the data are well
reproduced in our results.
Rescattering (FSI) effects are particularly important in
this respect: their parameters are derived from experimental
data on nearby resonances. The masses of these being unequal, flavour
SU(3) breaking is induced through the difference in the phase shifts for each
isospin channel.

W-exchange/annihilation contributions are substantial in many cases.
The danger of
getting too big decay rates for $D_s$ Cabibbo-favoured decays has been
avoided in our model imposing chiral symmetry requirements.

Moreover, the rather large final state phase shifts and ``penguin''
operator contributions lead us to envisage CP violating asymmetries
larger than $10^{-3}$ in some decay channels.
\vfill\eject

\begintable
$f_i$
| B.R.$_{exp}(D^0 \rightarrow f_{i})$ | B.R.$_{th}(D^0
\rightarrow f_{i})$ \cr
$K^{-} \pi^+ $  |$4.01 \pm 0.14$      |  $4.03$ \cr
$K_{S} \pi^{0}$ |$1.02 \pm 0.13$      | $0.78$ \cr
$K_{L} \pi^{0}$ |$-$                  | $0.57$ \cr
$K_{S}\; \eta$  |$0.34 \pm 0.06$      | $0.46$ \cr
$K_{L}\; \eta$  |$-$                  | $0.34$ \cr
$K_{S}\; \eta^{\prime}$ | $0.83 \pm 0.15$   | $0.84$ \cr
$K_{L}\; \eta^{\prime}$ | $ - $             | $0.67$ \cr
$\overline{K}^{*0} \pi^0 $ |$3.0 \pm 0.4 $  | $3.49$ \cr
$K_{S}\; \rho^0  $ | $0.55 \pm 0.09 $       | $0.49$  \cr
$K_{L}\; \rho^0  $ | $ - $                  | $0.39$  \cr
$K^{*-} \pi^+ $    |$4.9 \pm 0.6 $   | $4.69$  \cr
$K^{-} \rho^+  $   |$10.4 \pm 1.3 $  | $11.19$  \cr
$\overline{K}^{*0} \eta $ |$ 1.9 \pm 0.5 $ | $0.51$ \cr
$\overline{K}^{*0} \eta^{\prime} $ | $ < 0.11 $ | $ 0.005 $ \cr
$K_{S}\; \omega  $ |$1.0 \pm 0.2 $  | $1.12$ \cr
$K_{L}\; \omega  $ |$-$  | $1.04$ \cr
$K_{S}\; \phi  $   | $0.415 \pm 0.060 $  | $0.42$ \cr
$K_{L}\; \phi  $   | $ - $  | $0.48$
\endtable
\vskip 1 cm
\centerline{TABLE 1}
\vskip 0.5 cm
\noindent
Decay branching ratios in percent for Cabibbo allowed
two-body $D^0$ nonleptonic decays.
\par\noindent
In the first column the experimental data are reported
(upper bounds are 90\% c.l.) \rPDG,
 in the second
column the theoretical values obtained in the best fit.
\par\noindent
See text for further explanations.
\vfill\eject

\vskip 4 cm
\begintable
$f_i$
| B.R.$_{exp}(D^{+} \rightarrow f_{i})$ | B.R.$_{th}(D^{+}
\rightarrow f_{i})$  \cr
$K_{S} \pi^{+}$            |$1.37 \pm 0.15$  | $1.08$  \cr
$K_{L} \pi^{+}$            |$ - $  | $1.43$  \cr
$\overline{K}^{*0} \pi^+ $ |$2.2 \pm 0.4 $ | $0.64$ \cr
$K_{S} \rho^+  $           |$3.30 \pm 1.25 $  | $5.28$ \cr
$K_{L} \rho^+  $           |$ - $    | $6.49$ \cr
$\pi^+ \pi^0 $             | $0.25 \pm 0.07 $ | $0.17$ \cr
$\pi^+ \eta $              | $0.75 \pm 0.25 $ | $0.36$ \cr
$\pi^+ \eta^{\prime}$      | $< 0.9 $ | $0.79$ \cr
$\overline{K}^{0} K^+ $    | $0.78 \pm 0.17$ | $0.86$ \cr
$\rho^0 \pi^+ $            | $< 0.14$  | $0.17$ \cr
$\rho^+ \pi^0 $            | $ - $     | $0.37$ \cr
$\rho^+ \eta $             | $< 1.2$   | $0.0002$ \cr
$\rho^+ \eta^{\prime} $    | $< 1.5$ | $0.13$ \cr
$\omega \pi^+ $            | $< 0.7$ | $0.035$ \cr
$\phi \pi^+ $              | $0.67 \pm 0.08$ | $0.59$ \cr
$\overline{K}^0 K^{*+} $   | $ - $ | $ 1.70 $ \cr
$\overline{K}^{*0} K^+ $   | $0.51 \pm 0.10 $ | $0.25$
\endtable
\vskip 1 cm
\centerline{TABLE 2}
\vskip 0.5 cm
\noindent
Decay branching ratios in percent for Cabibbo allowed and first-forbidden
two-body $D^+$ nonleptonic decays.
\par\noindent
In the first column the experimental data are reported
(upper bounds are 90\% c.l.)\rPDG,
 in the second
column the theoretical rates obtained in the best fit.
\par\noindent
See text for further explanations.
\vfill\eject

\begintable
$f_i$
| B.R.$_{exp}(D_s^{+} \rightarrow f_{i})$ | B.R.$_{th}(D_s^{+}
\rightarrow f_{i})$  \cr
$K_{S} K^{+}$          |$1.75 \pm 0.35$  | $2.53$  \cr
$K_{L} K^{+}$          |$ - $  | $2.26$  \cr
$\pi^+ \eta $          | $1.90 \pm 0.40 $ | $1.33$ \cr
$\pi^+ \eta^{\prime} $ | $4.7 \pm 1.4$ | $5.89$ \cr
$\rho^+ \eta $         | $10.0 \pm 2.2$ | $9.49$ \cr
$\rho^+ \eta^{\prime} $| $12.0 \pm 3.0$ | $2.61$ \cr
$\overline{K}^{*0} K^+$|$3.3 \pm 0.5 $  | $3.86$ \cr
$K_{S} K^{*+}  $       |$2.1 \pm 0.5 $ | $1.44$ \cr
$K_{L} K^{*+}  $       |$ - $ | $1.93$ \cr
$\phi \pi^+ $          | $3.5 \pm 0.4$  | $2.89$ \cr
$\omega \pi^+ $        | $< 1.7$  | $0.0$ \cr
$\rho^0 \pi^+ $        |$< 0.28$  | $0.080$ \cr
$\rho^+ \pi^0 $        |$-$ | $0.080$ \cr
$K^+ \pi^0$            |$-$ | $0.16$  \cr
$K^+ \eta $            | $-$ | $0.27$ \cr
$K^+ \eta^{\prime} $   | $-$ | $0.52$ \cr
$K^0 \pi^+$            |$< 0.7$  | $0.43$  \cr
$K^{*+} \pi^0$         |$-$ | $0.029$ \cr
$K^+ \rho^0 $          | $-$ | $0.24$ \cr
$K^{*+} \eta $         |  $-$ | $0.024$ \cr
$K^{*+} \eta^{\prime} $| $-$  | $0.024$ \cr
$K^+ \omega$           | $-$  | $0.072$ \cr
$K^+ \phi$             | $< 0.25$ | $0.015$ \cr
$K^{*0} \pi^+ $        |$-$ | $0.33$ \cr
$K^0 \rho^+$           |$-$ | $1.95$
\endtable
\vskip 0.1 cm
\centerline{TABLE 3}
\vskip 0.02 cm
\noindent
\centerline{Same as Table 2 for $D_s^+$ nonleptonic decays.}
\vfill\eject

\begintable
$f_i$
| B.R.$_{exp}(D^0 \rightarrow f_{i})$ | B.R.$_{th}(D^0
\rightarrow f_{i})$ \cr
$\pi^0 \eta$               |$-$ | $0.058$ \cr
$\pi^0 \eta^{\prime}$      | $-$ | $0.17$ \cr
$\eta \eta$                |$-$ | $0.10$ \cr
$\eta \eta^{\prime}$       |$-$ | $0.22$ \cr
$\pi^0 \pi^0$              |$0.088 \pm 0.023$ | $0.116$ \cr
$\pi^+ \pi^-$              |$0.159 \pm 0.012$ | $0.159$ \cr
$K^+ K^-$                  |$0.454 \pm 0.029$ | $0.456$ \cr
$K^0 \KB$                  |$0.11 \pm 0.04$   | $0.093$ \cr
$\omega \pi^0$             | $-$   | $0.008$ \cr
$\rho^0 \eta$              | $-$   | $0.024$ \cr
$\rho^0 \eta^{\prime}$     | $-$   | $0.010$ \cr
$\omega \eta$              | $-$   | $0.19$ \cr
$\omega \eta^{\prime}$     | $-$   | $0.0001$ \cr
$\phi \pi^0$               | $-$   | $0.11$ \cr
$\phi \eta$                | $-$   | $0.057$ \cr
$K^{*0} \KB$               | $< 0.08$ | $0.099$ \cr
$\overline{K}^{*0} K^0 $   | $< 0.15$ | $0.099$  \cr
$K^{*+} K^-$               | $0.34 \pm 0.08$ | $0.45$ \cr
$K^{*-} K^+$               | $0.18 \pm 0.10$ | $0.28$ \cr
$\rho^+ \pi^-$             | $-$ | $0.82$ \cr
$\rho^- \pi^+$             | $-$ | $0.65$ \cr
$\rho^0 \pi^0$             | $-$ | $0.17$
\endtable
\vskip 0.3 cm
\centerline{TABLE 4}
\vskip 0.2 cm
\noindent
Decay branching ratios in percent for Cabibbo first-forbidden
two-body $D^0$ nonleptonic decays.
See the caption of Table 2 and text for details.
\vfill\eject

\vskip 4 cm
\begintable
decay channel |
${\displaystyle 10^3 \times a_{CP}}$ ||
decay channel |
${\displaystyle 10^3 \times a_{CP}}$ \cr

$D^+ \to \rho^0 \pi^+$ | $ -1.17 \pm 0.68 $ ||
$D^+ \to \KB K^+ $ | $ -0.51 \pm 0.30$ \cr
$D^+ \to \rho^+ \pi^0$ | $ +1.28 \pm 0.74$ ||
$D^0 \to \pi^0 \eta$ | $+ 1.43 \pm 0.83$ \cr
$D^0 \to K^{*0} \KB$ | $+0.67 \pm 0.39$ ||
$D^0 \to \pi^0 \eta^{\prime}$ | $- 0.98 \pm 0.57$ \cr
$D^0 \to \overline{K}^{*0} K^0 $ | $+0.67 \pm 0.39$ ||
$D^0 \to \eta \eta$ | $+ 0.50 \pm 0.29$  \cr
$D^0 \to K^{*+} K^-$ | $-0.038 \pm 0.022$ ||
$D^0 \to \eta \eta^{\prime}$ | $+0.28 \pm 0.16$ \cr
$D^0 \to K^{*-} K^+$ | $- 0.16 \pm 0.09$ ||
$D^0 \to \pi^0 \pi^0$ | $-0.54 \pm 0.31$  \cr
$D^0 \to \rho^+ \pi^-$ | $-0.37 \pm 0.22$ ||
$D^0 \to \pi^+ \pi^-$ | $+ 0.02 \pm 0.01$ \cr
$D^0 \to \rho^- \pi^+$ | $+0.36 \pm 0.21$ ||
$D^0 \to K^+ K^-$ | $+ 0.13 \pm 0.8$ \cr
 | || $D^0 \to K^0 \KB$ | $-0.28 \pm 0.16$
\endtable
\vskip 1 cm
\centerline{TABLE 5}
\vskip 0.5 cm
\noindent
CP-violating decay asymmetries for some $D^+$ and $D^0$ Cabibbo
forbidden decays. See text for explanation.
\vfill\eject
\listrefs
\bye